\documentclass[%
 reprint,
 amsmath,amssymb,
 aps,
]{revtex4-2}

\usepackage{graphicx}
\usepackage{dcolumn}
\usepackage{bm}
\usepackage{color,xcolor}

\begin{document}

\preprint{APS/123-QED}

\title{Geometric curvature effect on suppressing the Ion-Temperature-Gradient mode near the magnetic axis}

\author{Tiannan Wu}
\author{Shaojie Wang}%
 \email{correspongding author: wangsj@ustc.edu.cn}
\affiliation{%
 Department of Engineering and Applied Physics, University of Science and Technology of China, Hefei 230026, China
}%




\date{\today}

\begin{abstract}
Global gyrokinetic simulation of the ion temperature gradient mode shows that the radial electric field ($E_r$) well upshifts the critical temperature gradient near the magnetic axis, in the weak but not in the strong magnetic shear configuration. The geometric curvature effect significantly influences the $\bm{E}\times\bm{B}$ shear and the wave number near the axis, so that the $E_r$ well suppresses the high-$n$ modes but has little effect on the low-$n$ modes, which are suppressed by the weak magnetic shear effect. This new finding unravels the formation mechanism of the internal transport barrier in the weak central magnetic shear discharges. 

\end{abstract}

\maketitle



\noindent\textbf{(I) Introduction. }
When the external heating power exceeds a critical value, a sudden transition of the fusion plasma from the low-confinement to the high-confinement state has been widely observed \cite{Wagner1982PRL, Wagner2007PPCF, Koide1994PRL, Burrell1998PPCF}, with the drift-wave turbulence, such as the Ion-Temperature-Gradient-driven (ITG) mode, suppressed, and a steep pressure gradient is established in a narrow layer known as the edge transport barrier \cite{Wagner1982PRL, Wagner2007PPCF} or the Internal Transport Barrier (ITB) \cite{ Koide1994PRL, Burrell1998PPCF, Ida2018PPCF, Connor2004NF}. Transport barriers have also been observed in many other systems, such as the atmospheric \cite{McIntyre1989JATP}, the oncological \cite{nizzero2018TC}, the ecophysiological \cite{Schreiber2010TPS}, the biomolecular \cite{Moran2024NRN}, and the geophysical system \cite{Olascoaga2006GRL}. Therefore, understanding the formation mechanism of the transport barriers is of broad interest. 

The ITB plays an important role in the success of the upcoming International Thermonuclear Experimental Reactor (ITER). The most promising ITB for ITER \cite{Doyle2007NF} is the one forming in a weak central magnetic shear configuration \cite{Koide1994PRL,  Koide1996PPCF, Burrell1998PPCF, Retting1998POP,  Yu2016NF, Li2022PRL}, because this magnetic configuration is reliable to control in the hybrid operation scenario \cite{Doyle2007NF, Fasoli2023PRL, Yu2016NF, Li2022PRL}. During the formation of this ITB, the ITG turbulence \cite{Biglari1989POP, Connor1994PPCF, Horton1999RMP} is suppressed and a steep ion temperature gradient emerges near the magnetic axis \cite{Koide1994PRL, Koide1996PPCF, Burrell1998PPCF, Retting1998POP}.

It is well-known that the ITG mode can be stabilized by the $\bm{E}\times\bm{B}$ shear effect \cite {Biglari1990PFB,Hahm1994POP,Terry2000RMP}. The first-principle gyrokinetic (GK) simulations \cite{Kinsey2005POP} have shown that the ITG turbulence is quenched if the $\bm{E}\times\bm{B}$ shear rate \cite{Hahm1994POP,Waltz1999POP} generated by the radial electric field ($E_r$) well is larger than the maximum linear growth rate. The $\bm{E}\times\bm{B}$ shear can be generated by the zonal flow \cite{Lin1998SCIENCE}, which is nonlinearly excited by the ITG turbulence, or generated by the mean flow \cite{Schmitz2013PRL}, which satisfies the ion radial force balance equation. One of the most important contribution to the mean $E_r$ is from the radial ion pressure gradient, which was found to play a key role in reducing the ITG turbulence \cite{Schmitz2013PRL} in the H-mode plasma. GK simulations have shown that the nonlinearly excited zonal flows substantially reduce the ITG turbulence \cite{Lin1998SCIENCE}, and this leads to the Dimits upshift \cite{Dimits2000POP} of the critical temperature gradient, at which the ITG mode is marginally stable.

The ITG mode can also be stabilized by the weak magnetic shear ($s$) effect. Early theories \cite{Kadomtsev1970Book, Wesson1997Book} showed that the magnetic shear has a stabilizing effect on the drift waves in the strong shear regime. However, in the weak magnetic shear regime, theories \cite{Romanelli1993POP, Connor2004PPCF} and numerical results \cite{Dong1992PF, Gorler2016POP} have shown that the growth rates of the ITG modes decreases with the magnetic shear when $s<s_c$. 

However, these theories can not be directly applied to understand the ITG instability near the magnetic axis, which is crucially important to understand the ITB formation mechanism, because the geometric curvature effects, which are related to the large factor of $1/r$, make the usual ballooning representation \cite{Connor1978PRL} break down near the axis; here $r$ is the minor radius of the torus. The geometric curvature near the axis also causes difficulties in numerical simulation \cite{Wu2024Arxiv, Jolliet2007CPC}. Recently, global GK codes, GT5D \cite{Idomura2008CPC}, ORB5 \cite{Jolliet2007CPC}, GKNET \cite{Imadera2023PPCF}, and NLT \cite{Wu2024Arxiv} have been developed to include the magnetic axis. Nonlinear GK simulations including the magnetic axis have been carried out to investigate the ITB formation dynamics \cite{Imadera2023PPCF,Wang2024PRL}. However,  
until recently, GK simulations \cite{Qin2018PFR, Xu2022POP} and gyrofluid simulation \cite{Ko2024NF} of the linear ITG mode are still carried out away from the axis. Therefore, it is of significant interest to investigate the instability of ITG mode near the magnetic axis. 

Here, we report for the first time that the mean $\bm{E}\times\bm{B}$ shear of a $E_r$ well significantly upshift the linear temperature gradient in the weak magnetic shear configuration near the magnetic axis, and this upshift vanishes in the strong magnetic shear configuration; this is consistent with the experimental observations that the ITB usually appears near the axis in the weak magnetic shear configuration. This linear upshift is different from the Dimits nonlinear upshift \cite{Dimits2000POP}. It is found that the geometric curvature effect significantly modifies the $\bm{E}\times\bm{B}$ shear of the near-axis $E_r$ well, which leads to that the $\bm{E}\times\bm{B}$ shear suppresses the ITG modes with high toroidal mode number $n$, but hardly affects the low-$n$ modes, which can be stabilized by the effects of weak central magnetic shear.

\noindent\textbf{(II) Method and Setup. }
We carry out the linear global GK simulation of the ITG mode near the magnetic axis with adiabatic electrons and kinetic ions by using the NLT code \cite{Ye2016JCP, Xu2017POP}, which solves the GK equation by using the numerical Lie transform method \cite{Wang2012POP, Wang2013PRE, Wang2013POP}. At the magnetic axis, the perturbed distribution function and the electrostatic potential are solved by using the mean value theorem, 
\begin{align}
    g(0) = \frac{4}{3N_{\theta}}\sum_{k=1}^{N_{\theta}}g(\Delta r,\theta_{k})-\frac{1}{3N_{\theta}}\sum_{k=1}^{N_{\theta}}g(2\Delta r,\theta_{k}),
\end{align}
which indicates that the value of a scalar function at the polar axis, $g(0)$, can be predicted by the average value of $g(\Delta r,\theta_{k})$ and $g(2\Delta r,\theta_{k})$, with $\Delta r$ the interval of the radial grid points and $\theta_{k}$ the poloidal grid points. The mean value theorem ensures the continuity of the physical scalar quantities near the axis \cite{Wu2024Arxiv}.  

\begin{figure}[hbtp]
\includegraphics[width=0.45\textwidth]{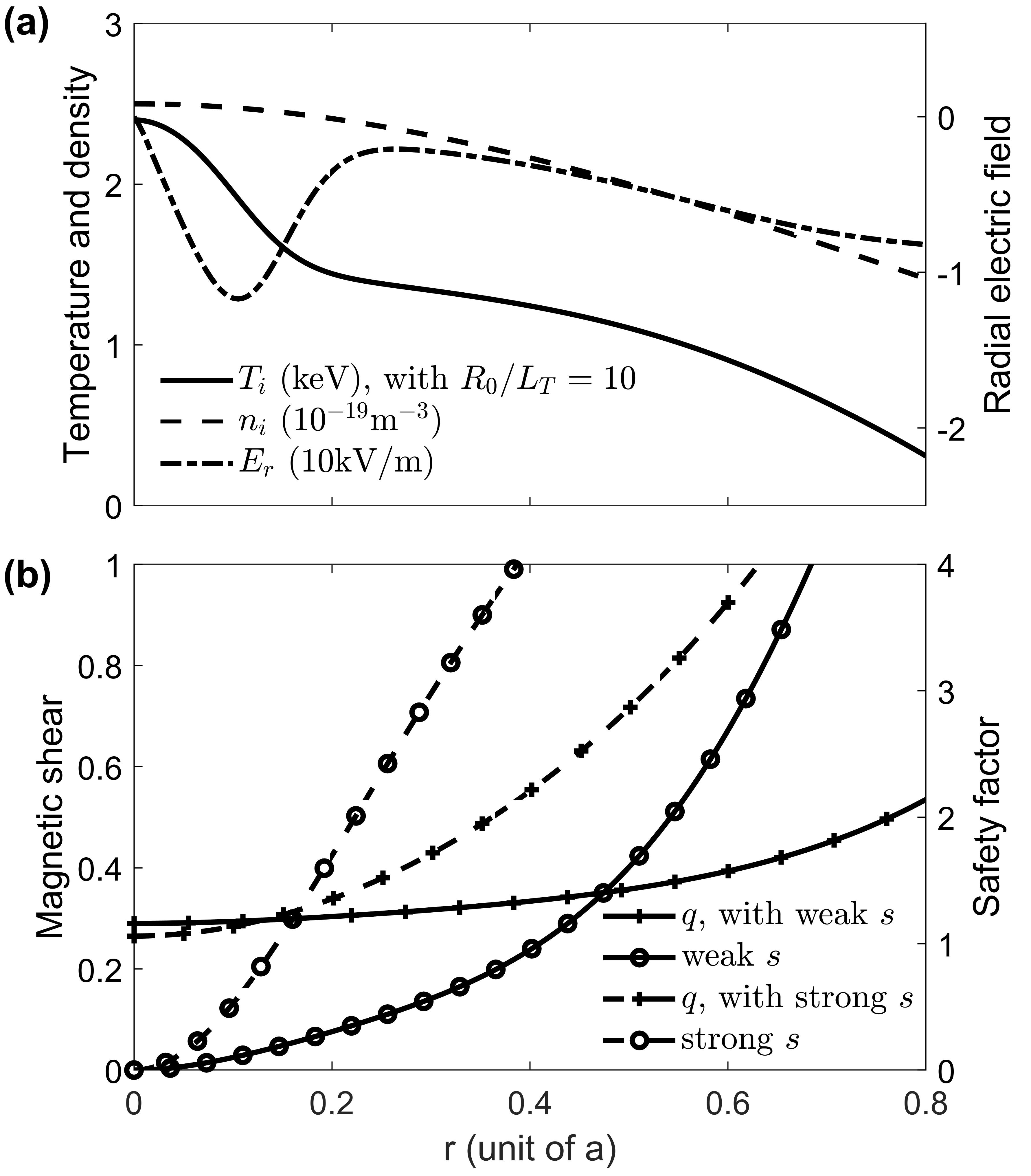}
\caption{\label{FigEqprofile} Equilibrium profiles. }
\end{figure}

The main parameters in our simulation are chosen to model a DIII-D-like deuterium plasma. The major and minor radius of the torus are $R/a=1.67\mathrm{m}/0.67\mathrm{m}$; the magnetic field at the magnetic axis is $B_0=2.1\mathrm{T}$. Equilibrium profiles of ion temperature $T_i$, ion density $n_i$, safety factor $q$ with weak and strong magnetic shear ($s$), and the $E_r$ well are shown in Fig. \ref{FigEqprofile}. $T_{e}=T_{i}$ is assumed. The equilibrium profiles are chosen to model the plasma with the ITB \cite{Burrell1998PPCF,Challis2002PPCF}. The gradient of the temperature profile is peaked at $r=0.1a$. The fitting formula of the temperature profile is given by $T_i(r) / T_{0, i}=\alpha_T\left[1-\tanh (r^2/\Delta_T^2)\right]+0.50-0.58 r^2+0.27 r^4-0.55 r^6$, with $\Delta_T=0.14$, $T_{0,i}=3\mathrm{keV}$. $\alpha_{T}$ is a parameter to control the maximum temperature gradient $R_0/L_{T}$, e.g., $R_0/L_{T}=10$ with $\alpha_{T}=0.3$. Here $E_r$ is chosen to balance the ion pressure gradient, which is consistent with the nonlinear simulation results \cite{Garbet2001POP,Wang2024PRL}; we have verified that the main conclusion of this paper is not changed by reducing $E_r$ with a factor of 0.8 or reversing the sign of $E_r$. The simulation domain is $r/a\in[0,0.8]$, $\theta\in[-\pi,\pi]$, $\alpha\in[0,2\pi/n]$, $v_{\parallel}/c_{s}\in[-4.3,4.3]$, $\mu B_{0}/T_{0,i}\in[0,4.3^2/2]$; here, $\theta$ is the poloidal angle, $\alpha=q\theta-\zeta$, with $\zeta$ the toroidal angle. $v_{\parallel}$ and $\mu$ are parallel velocity and magnetic moment, respectively. $m_{i} c_{s}^{2}=T_{0,i}$, with $m_i$ the ion mass. Grid numbers are $N_{r}\times N_{\theta}\times N_{\alpha}\times N_{v_{\parallel}}\times N_{\mu}=178\times 16\times 16\times 64\times 16$. 

\begin{figure}[htbp]
\includegraphics[width=0.45\textwidth]{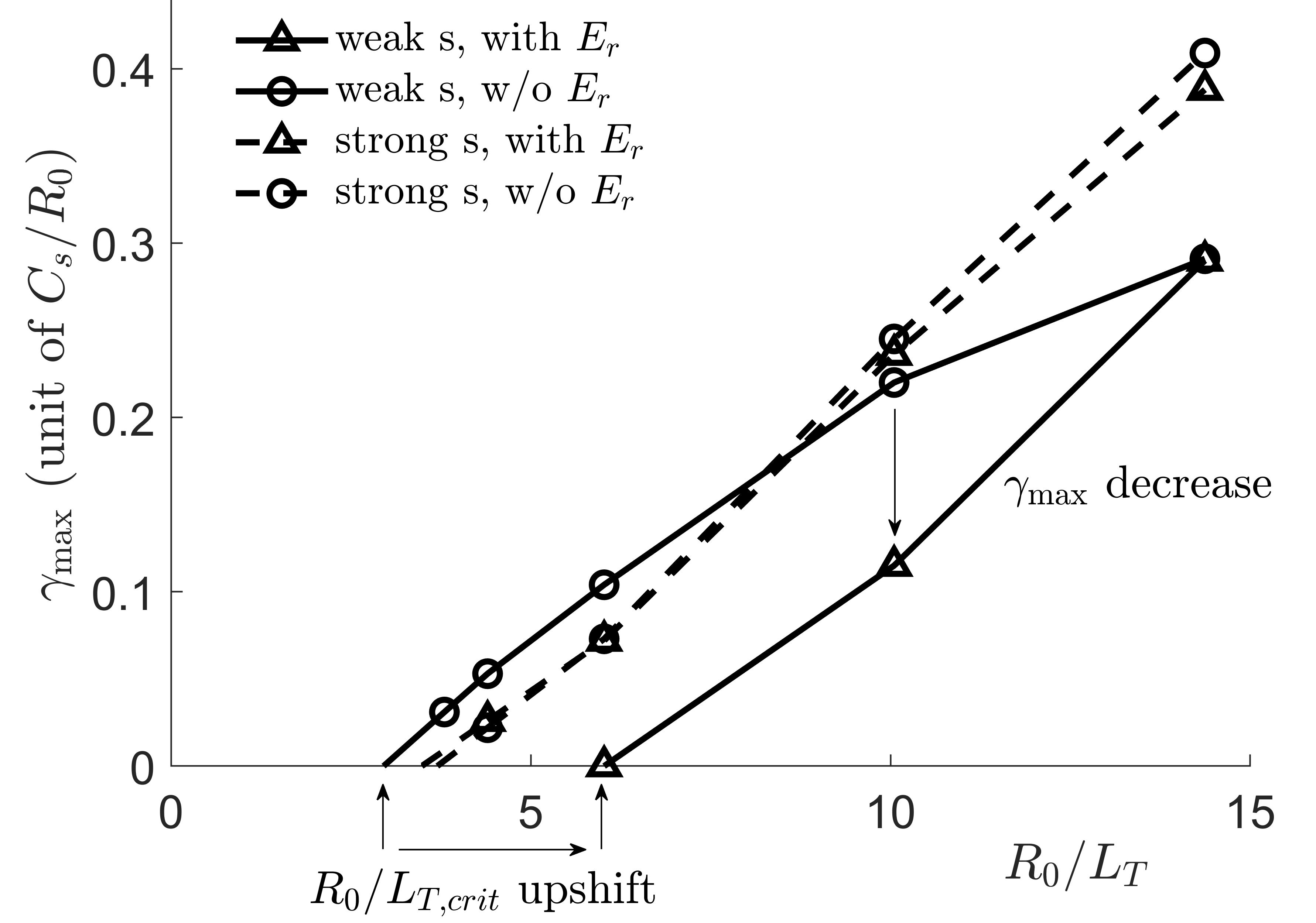}
\caption{\label{FigUpshift} Effects of the $E_r$ well and the magnetic shear on the instability of ITG mode near the magnetic axis.}
\end{figure}

\noindent\textbf{(III) Results and Discussions. } 
The linear ITG mode simulations are performed for different temperature gradients in four cases: with weak and strong magnetic shear, with and without $E_{r}$. The maximum linear growth rates $(\gamma_{max})$ of the modes are shown in Fig. \ref{FigUpshift}. In the weak magnetic shear configuration, $E_{r}$ significantly upshifts the linear critical temperature gradient $(R_{0}/L_{T,crit})$, where $L_{T,crit}$ is the temperature gradient scale length under which $\gamma_{max}=0$, or significantly reduces $\gamma_{max}$. However, in the strong magnetic shear configuration, $E_{r}$ hardly upshifts the linear critical temperature gradient or reduces $\gamma_{max}$. We emphasize that the upshift found here is a linear upshift caused by the mean $E_r$, which is different from the Dimits upshift, which is a nonlinear upshift caused by the zonal $E_r$. More importantly, this upshift happens in the weak magnetic configuration, but it does not happen in the strong magnetic shear configuration. A magnetic shear scan indicates that the critical value of the magnetic shear $s_c$, below which the upshift can be observed, is $s_c\approx 0.1$. This new finding is consistent with the ITB experiments, in which the near-magnetic-axis ITB is usually observed in weak central magnetic shear discharges, often accompanied by the sawtooth oscillations \cite{Yu2016NF,Joffrin2002PPCF,Chung2021NF}.  

\begin{figure}[htbp]
\includegraphics[width=0.45\textwidth]{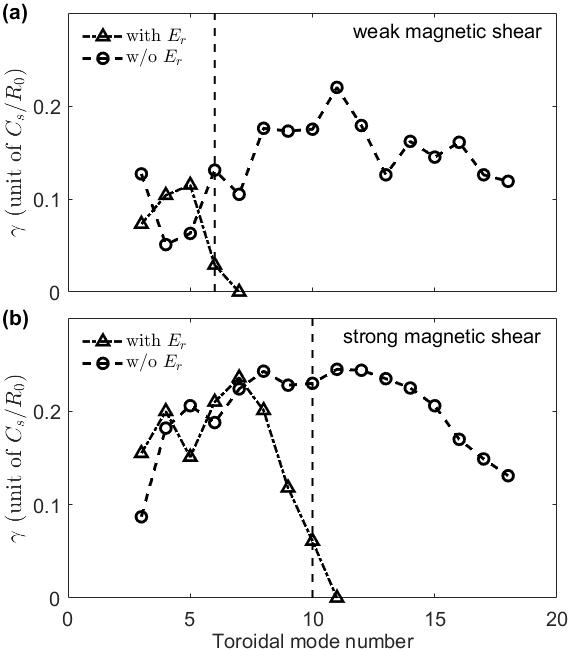}
\caption{\label{FigScanN} Effects of $E_r$ on the growth rates of ITG modes for different toroidal mode numbers, with $R_0/L_T=10$ fixed. (a) the weak and (b) the strong magnetic shear configurations. The high-$n$ ($n>n_c$) modes are stabilized by $E_{r}$. The critical toroidal mode numbers in the weak and strong magnetic shear configurations are respectively $n_{c}=6,10$.}
\end{figure}

To understand the finding given in Fig. \ref{FigUpshift}, we scan the toroidal mode number $(n)$ for the 4 cases, with $R_0/L_T=10$; the results are shown in Fig. \ref{FigScanN}, which suggests that, in both  magnetic shear configurations, there is a critical toroidal mode number $(n_c)$ above which the ITG mode is stabilized by $E_r$, although $n_c$ is different in the weak and the strong magnetic shear configurations. The maximum linear growth rate of the low-$n$ ($n<n_c$) modes, $(\gamma_{low-n})$, hardly decreases with $E_r$, as is summarized in TABLE \ref{TabGammaNc}. This indicates that the ITG instability for the cases with $E_r$ is determined by the instability of the low-$n$ modes. For the cases without $E_r$, $\gamma_{low-n}<\gamma_{high-n}$ in the weak magnetic shear configuration, while  $\gamma_{low-n} \approx \gamma_{high-n}$ in the strong magnetic shear configuration. Therefore, after the high-$n$ modes are stabilized by $E_{r}$, $\gamma_{max}$ decreases significantly only in the weak magnetic shear configuration; this improves our understanding of  Fig. \ref{FigUpshift}. 

\begin{table}[htbp]
 \renewcommand{\arraystretch}{1.5}
\begin{ruledtabular}
\begin{tabular}{lcccc}
&\multicolumn{2}{c}{weak $s$}&\multicolumn{2}{c}{strong $s$}\\
 &$\gamma_{low-n}$&$\gamma_{high-n}$&
 $\gamma_{low-n}$ &$\gamma_{high-n}$\\ 
 \hline
w/o $E_{r}$ & 0.127 & 0.220 & 0.243 & 0.245\\
with $E_{r}$& 0.115 & stabilized & 0.236 & stabilized\\
\end{tabular}
\end{ruledtabular}
\caption{\label{TabGammaNc}%
Effects of $E_r$ on the linear instability of low-$n$ and high-$n$ modes, with $R_0/L_T=10$. 
}
\end{table}

\begin{figure} [htbp]
\includegraphics[width=0.45\textwidth]{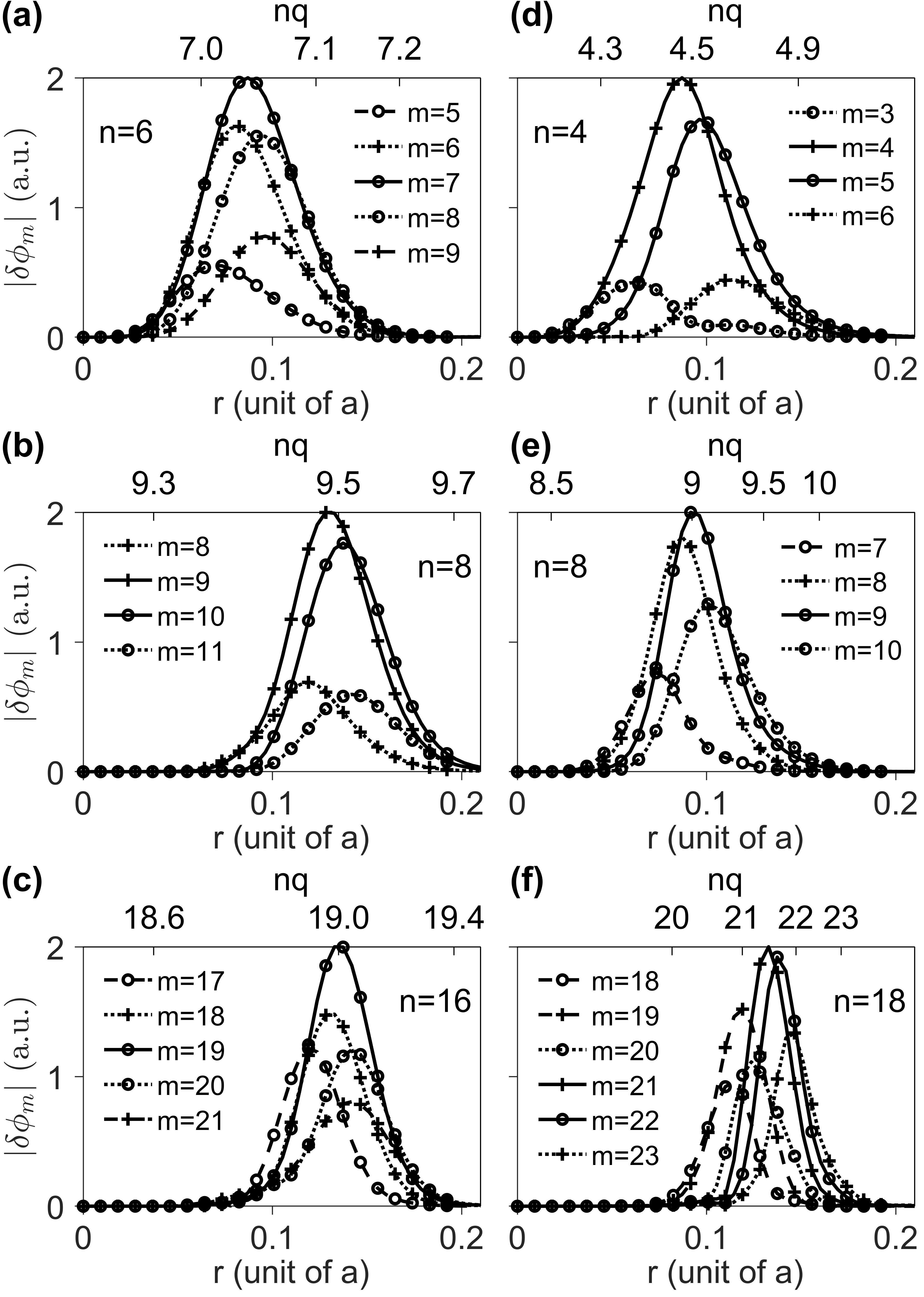}
\caption{\label{FigPhim} Poloidal Fourier components of 3 types of unconventional mode near the magnetic axis. (a, b, c) weak magnetic shear case. (d, e, f) strong magnetic shear case. (b, d) Type 1, without rational surfaces. (a, c, e) Type 2, with one rational surface. (f) Type 3, with more than 1 rational surfaces.}
\end{figure}

Fig. \ref{FigScanN} and TABLE \ref{TabGammaNc} indicate that 
$\gamma_{low-n}$ decreases with the magnetic shear when $s<s_c$. This is consistent with the previous results that ITG modes are more stable in the weaker magnetic shear configuration if the magnetic shear is sufficiently weak ($s<s_c$) \cite{Romanelli1993POP, Connor2004PPCF, Dong1992PF, Kinsey2006POP}.  Although $s_c\approx 0.1$ is found here for the near-magnetic-axis modes, it should be pointed out that $s_c\approx 0.5$ for the conventional modes away from the magnetic axis \cite{Dong1992PF, Kinsey2006POP}, which has been verified by using the NLT code. The unconventional radial structure of the ITG mode near the magnetic axis, which is shown in Fig. \ref{FigPhim}, may account for this difference. Fig. \ref{FigPhim} shows that the width of the ITG mode, except for the high-$n$ mode in the strong magnetic shear configuration shown in  Fig. \ref{FigPhim} (f), is smaller than the distance between the adjacent mode rational surfaces. Figs. \ref{FigPhim}(b, d) show Type 1 mode which has not a rational surface, and this type of mode consists of two dominant poloidal Fourier components, $m$ and $m+1$, which are coupled near the surface with $nq=m+1/2$; Figs. \ref{FigPhim}(a, c, e) shows Type 2 mode which has one rational surface, and this type of mode consists of three dominant poloidal Fourier components, $m$ and $m\pm 1$, which are coupled near the surface with $nq=m$. Fig. \ref{FigPhim}(f) shows Type 3 mode, which has more than 1 rational surfaces; the dominant poloidal Fourier component ($\delta\phi_{m}$) is not at the corresponding rational surface, where $nq=m$. Note that Type 1 and Type 2 modes are consistent with Ref. \onlinecite{Connor2004PPCF}, while Type 3 is consistent with Ref. \onlinecite{Xie2015POP}.

Fig. \ref{FigScanN} and TABLE \ref{TabGammaNc} suggest that the results with $E_r$ shown in Fig. \ref{FigUpshift} is determined by the behavior of the low-$n$ modes. This has been verified by plotting $\gamma_{low-n}$ vs. $R_0/L_T$ for both magnetic shear configurations with $E_r$, and the result agrees well with Fig. \ref{FigUpshift}. 

\begin{figure} [htbp]
\includegraphics[width=0.45\textwidth]{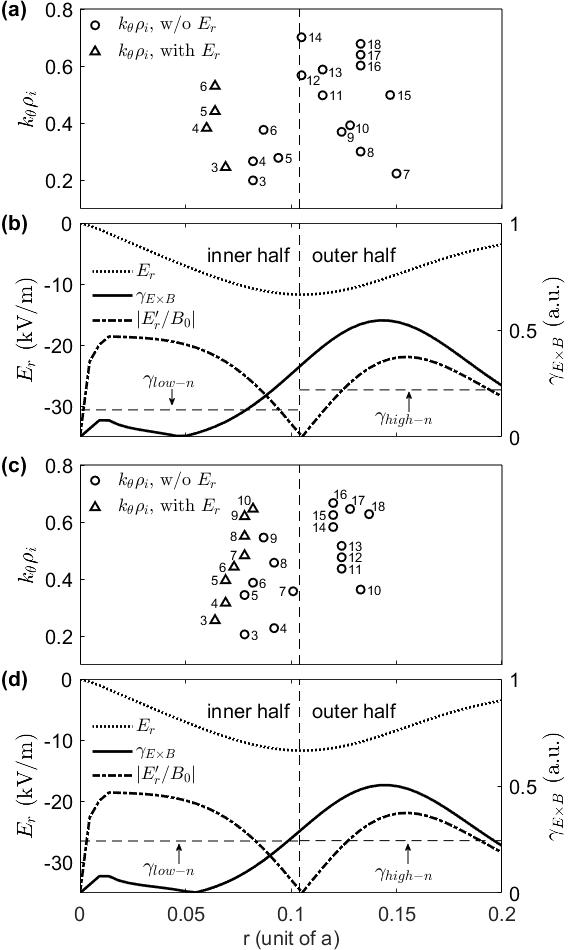}
\caption{\label{FigGeoEff} Geometric curvature effects on the $\bm{E} \times \bm{B}$ shear and $k_{\theta}\rho_i$. (a, b) the weak and (c, d) the strong magnetic shear configurations. (a, c) The radial position of the envolpe and $k_{\theta}\rho_i$ for each mode (maker); the number near the marker denotes the toroidal mode number. (b, d) $\bm{E} \times \bm{B}$ shear rate.}
\end{figure}

To understand Fig. \ref{FigScanN}, we need to reveal the key physics of the near-magnetic-axis ITG mode, to explain why the $E_r$ well, which is usually effective in stabilizeing the conventional ITG mode away from the axis, is effective to stabilize the high-$n$ modes but ineffective to stabilize the low-$n$ modes near the axis. To proceed, we plot in Fig. \ref{FigGeoEff} the radial position of the envolpe and $k_{\theta}\rho_i$ for different modes. 
\begin{equation}\label{eq:krho}
k_{\theta}\rho_{i}\approx \frac{nq(r_p)}{r_p}\rho_i, 
\end{equation}
with $\rho_i=m_i c_s / e_i B_0$; here, $r_p$ is the radial position of the ITG mode envelope, and $e_i$ is the ion charge. The $\bm{E} \times \bm{B}$ shear rate plotted here is given by \cite{Waltz1999POP} 
\begin{align}\label{eq:EXB}
    \gamma_{E\times B} = \frac{1}{B_0}\left|E'_r-\frac{1-s}{r}E_{r}\right|,
\end{align}
with $E'_{r}=\mathrm{d}E_r/\mathrm{d}r$.  

Figs. \ref{FigGeoEff}(a, c) show that, without $E_{r}$, the low-$n$ and high-$n$ modes are respectively located in the inner half and outer half of the $E_r$ well; this is consistent with the fact that the most unstable ITG mode is the one with $k_{\theta}\rho_{i}\approx 0.4$ \cite{Dong1992PF, Gorler2016POP}. When approaching the axis, $r$ decreases, therefore the poloidal wave-vector ($k_{\theta}$) of the low-$n$ mode increases due to the geometric curvature effect, this is given by Eq. (\ref{eq:krho}). Figs. \ref{FigGeoEff}(b, d) show that the $\bm{E} \times \bm{B}$ shear rate is significantly reduced in the inner half but significantly enhanced in the outer half; this can be understood by checking Eq. (\ref{eq:EXB}). Note that $E'_r$ represents the contribution from the shear of the $E_r$ well, and $\frac{1-s}{r}E_{r}$, which is approximately $E_r/r$ here, represents the contribution from the geometric curvature effect; for the near-axis $E_r$ well, the $E'_r$ term is almost cancelled by the $E_r/r$ term in the inner half, while the $E'_{r}$ term is almost doubled by the $E_r/r$ term in the outer half.
Therefore, due to the geometric curvature effect, the low-$n$ modes locate in the inner half while the high-$n$ modes locate in the outer half. The high-$n$ modes are stabilized by the $\bm{E} \times \bm{B}$ shear, since the linear stabilization condition \cite{Waltz1999POP}, $\gamma_{\bm{E} \times \bm{B}}>\gamma_{max}$, is easier to be satisfied in the outer half of the $E_r$ well, due to the enhancement by the geometric curvature effect, as is clearly shown in Figs. \ref{FigGeoEff}(b, d).

\noindent\textbf{(IV) Summary. }
In conclusion, the global GK simulation shows that the mean $E_r$ well upshifts the linear critical temperature gradient of the ITG mode near the magnetic axis in the weak but not in the strong magnetic shear configuration. Due to the geometric curvature effect, the low-$n$ and the high-$n$ modes locate in the inner and outer halves of the $E_r$ well near the axis, respectively. The geometric curvature effect almost cancels and doubles the $E_r$ shear effect in the inner and outer halves of the well near the axis, respectively. Therefore, near the axis, the $E_r$ well suppresses the high-$n$ modes but has little effect on the low-$n$ modes, which can be suppressed by the weak magnetic shear effect. 

The new finding reported here suggests the following scenario of the ITB formation in the weak central magnetic shear discharges \cite{Koide1994PRL, Burrell1998PPCF, Retting1998POP, Yu2016NF, Li2022PRL}. When the central magnetic shear is lowered below the critical value, the low-$n$  modes are supressed near the axis by the weak magnetic shear effect; this reduces the ITG turbulence and steepens the ion temperature gradient near the axis; this steepening of the ion temperature gradient, through the ion radial force balance condition, establishes the $E_r$ well, which further suppresses the high-$n$ mdoes by the $\bm{E}\times\bm{B}$ shear enhanced by the geometric curvature effect. 
The geometric curvature effects reported here may also be of interest in the atomspheric physics near the pole of a planet. 

\begin{acknowledgments}
This work was supported by the National MCF Energy R\&D Program of China under Grant No. 2019YFE03060000, and The Strategic Priority Research Program of the Chinese Academy of Sciences under Grant No. XDB0790201. 
\end{acknowledgments}

\nocite{*}


\providecommand{\noopsort}[1]{}\providecommand{\singleletter}[1]{#1}%

\end{document}